\documentclass[showpacs,prl,preprintnumbers,amsmath,amssymb,twocolumn]{revtex4}
\usepackage[latin1]{inputenc}
\usepackage{graphicx}
\usepackage{dcolumn}
\usepackage{bm}

\begin{document}

\title{Controlled calculation of the thermal conductivity for a spinon Fermi surface coupled to a $U(1)$ gauge field}
\author{Hermann Freire$^{1,2}$}
\email{hfreire@mit.edu}
\email[]{hermann\_freire@ufg.br}
\affiliation{$^{1}$Instituto de Física, Universidade Federal de Goiás, 74.001-970, Goiânia-GO, Brazil}
\affiliation{$^{2}$Department of Physics, Massachusetts Institute of Technology, Cambridge, Massachusetts 02139, USA}

\date{\today}

\begin{abstract}
Motivated by recent transport measurements on the candidate spin-liquid phase of the organic triangular lattice insulator EtMe$_3$Sb[Pd(dmit)$_2$]$_2$, 
we perform a controlled calculation of the thermal conductivity
at intermediate temperatures in a spin liquid system where a spinon Fermi surface is coupled to a $U(1)$
gauge field. 
The present computation builds upon the double expansion approach developed by Mross \emph{et al.} [Phys. Rev. B \textbf{82}, 045121 (2010)]
for small $\epsilon=z_b -2$ (where $z_b$ is the dynamical critical exponent of the gauge field) and large number of fermionic species $N$.
Using the so-called memory matrix formalism that most crucially does not assume the existence of well-defined quasiparticles at low energies in the system, 
we calculate the temperature dependence of the thermal conductivity $\kappa$ of this model due to non-critical Umklapp
scattering of the spinons for a finite $N$ and small $\epsilon$. Then we discuss the physical implications of such theoretical result in connection with
the experimental data available in the literature.
\end{abstract}

\pacs{74.20.Mn, 74.20.-z, 71.10.Hf}

\maketitle

In the last few years, there has been immense experimental progress in unveiling the nature of promising
quantum spin liquid (QSL) phases displayed by some insulating materials featuring a two-dimensional (2D) Heisenberg triangular lattice, such as,
the organic compounds $\kappa$-(BEDT-TTF)$_{2}$-Cu$_{2}$(CN)$_{3}$ \cite{Saito,Kanoda} and EtMe$_3$Sb[Pd(dmit)$_2$]$_2$  
(or simply dmit-131) \cite{Yamashita,Itou,Watanabe}.
In both materials, while the antiferromagnetic exchange coupling is approximately $J\sim 250$ K, nuclear magnetic
resonance (NMR) measurements have reported no evidence of long-range magnetic order down to $T=20$ mK \cite{Saito,Itou}. Moreover,
in the case of dmit-131, despite the fact that it is clearly a charge insulator, a ``Fermi-liquid''-like low-temperature
behavior has been observed in both the static susceptibility which remains finite down to the lowest temperatures
measured \cite{Watanabe} and the specific heat which becomes linear in $T$ \cite{Yamashita}. This latter observation suggests gapless spin excitations 
present in the system. This interpretation is supported by transport measurements performed in Ref. \cite{Yamashita} where a linear 
dependence in $T$ was also reported in the thermal conductivity $\kappa$ at low temperatures and no thermal
Hall angle as a function of magnetic field $H$ was observed within error bars.

Given the above experimental results one may legitimately ask what is the QSL state which could be at work in the dmit-131 material?
One possible mean-field solution of the $t-J$ model on the triangular lattice is obtained by applying the slave-boson technique to
enforce no double occupancy of particles on a single lattice site. By including fluctuations around this state, one arrives
at an effective theory with a $U(1)$ emergent gauge field coupled to neutral spinons \cite{Motrunich,PALee2,Kim,Nave_Lee,Lee_Lee,PALee}. 
Recently, it has been argued \cite{SSLee2} that instanton effects
are irrelevant in the low-energy limit for this model due to the existence of a spinon (critical) Fermi surface and, as a result, a deconfined 
state may indeed occur. Therefore, in this work we pragmatically
take this QSL state as a possible hypothesis and examine in a controlled way what is the thermal conductivity
of this model at finite temperatures with the goal in mind of trying to compare this result with the experimental situation. In addition
to this, we mention here that 
another QSL state has also been proposed to describe the dmit-131 recently in the literature \cite{Kivelson}.

Following the approach introduced by Ref. \cite{Mross} which builds upon previous works by Refs. \cite{Nayak,SSLee,Metlitski}, 
our present model is described by an effective two-patch 
low-energy Euclidean action $S=S_{f}+S_{int}+S_{a}$ given by

\vspace{-0.2cm}

\begin{eqnarray}\label{action}
S_f&=&\int d^2 x\, d\tau \sum_{s\alpha}\overline{\psi}_{s\alpha}(\eta\,\partial_{\tau}-is v_x \partial_{x}- v_y \partial_{y}^{2})\psi_{s\alpha}\nonumber\\
S_{int}&=&\int d^2 x\, d\tau \sum_{s\alpha}
\frac{g_{s}}{\sqrt{N}} (1+e^{i\mathbf{K.x}})\,a\, \overline{\psi}_{s\alpha}\psi_{s\alpha}\nonumber\\
S_{a}&=&\int_{\mathbf{k},k_0} (\eta'\,k_{0}^2+|k_{y}|^{1+\epsilon})|a(\mathbf{k},k_{0})|^{2},
\end{eqnarray}

\noindent which includes the most singular kinematic regime in the system \cite{Polchinski,Altschuler}, 
i.e., two antipodal fermionic spinons $\psi_{s\alpha}$ for $s=\pm$ (see Fig. 1) interacting with
a gapless gauge field $a(\mathbf{x},\tau)$ and $\alpha=1,...,N$ is the flavor index. 
We have chosen the Fermi velocity to be along the $x$-direction where $v_x$ is the Fermi velocity and $v_y$
determines the curvature of the Fermi surface.
Unless otherwise stated, we shall set $v_x$ and $v_y$ to unity throughout this work.
We include here the possibility of Umklapp scattering in the present model, which conserves
the quasi-momentum of the spinons up to a reciprocal lattice vector $\mathbf{K}$.
This will provide a mechanism for momentum relaxation in the present system.
In line with some of the conclusions in Ref. \cite{Chubukov} (where the role of Umklapp scattering in a closely related problem
was thoroughly analyzed using a semiclassical Boltzmann-equation approach), the spinon Fermi surface must be large enough for this mechanism
to be effective at intermediate temperatures, since the ``gap'' between the Fermi surface and the edges
of the Brillouin zone would be small in this case.
A controlled expansion can be developed for small $\epsilon=z_b-2$, where $z_b$ 
is the dynamical critical exponent of the gauge field $a(\mathbf{x},\tau)$, and large number of fermionic species $N$.
For a $U(1)$ spin liquid, 
$g_{\pm}=\pm g$, where $g$ is the coupling constant. The renormalization group (RG) scaling
that one considers for this model is: $k_{0}\rightarrow b^{(1+\frac{g^2}{4\pi^2 N})}k_{0}$, $k_x\rightarrow b k_x$, $k_y\rightarrow b^{1/2} k_y$, 
$\psi_{s\alpha}(\mathbf{x},\tau)\rightarrow b^{(\frac{3}{4}+\frac{g^2}{8\pi^2 N})}\psi_{s\alpha}(\mathbf{x},\tau)$, 
$a(\mathbf{x},\tau)\rightarrow b^{(1-\frac{\epsilon}{4}+\frac{g^2}{8\pi^2 N})}a(\mathbf{x},\tau)$, 
and $g_{s}\rightarrow b^{(\frac{\epsilon}{4}-\frac{g^2}{8\pi^2 N})}g_{s}$. As a result of this scaling,
one finds straightforwardly the nontrivial fixed point in the infrared (IR) limit given by $g^{2}_{*}=2\pi^2 N\epsilon$ \cite{Nayak}.
The existence of this nontrivial fixed point which is accessible perturbatively for finite $N$ and
small $\epsilon$ will be central to our controlled calculation of the thermal conductivity for this model.

\begin{figure}[t]
 \centering
 \includegraphics[height=2.3in]{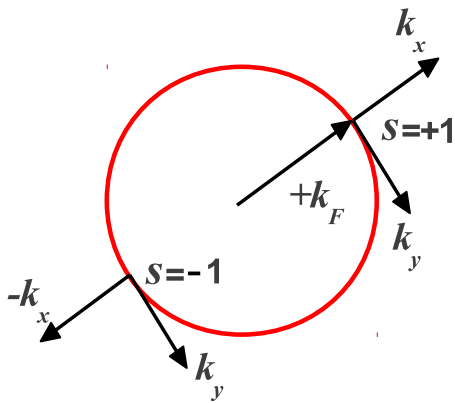}
 \caption{(Color online) The two-patch model considered in the present work involves
two types of fermion fields ($s=\pm$) with momenta close to two
antipodal patches on the spinon Fermi surface centered at $\pm k_F$ interacting with a gauge field. The
fermionic momenta are measured locally with respect to $\pm k_F$.
We define $k_x$ and $k_y$, respectively, as the normal and parallel components to the Fermi surface.}
\end{figure}

It is interesting to note that under this scaling, the parameter $\eta'$ should
scale as $\eta'\rightarrow b^{(\epsilon-3)/2}\eta'$ and therefore, for small $\epsilon$, it is always irrelevant 
in the RG sense. The parameter $\eta$ is dimensionless exactly at $z_b=2$ (i.e. marginal for this case) and
becomes the least irrelevant term for $2 < z_b\leq 3$. As a result, the ratio $(\eta'/\eta)$ always flows to zero for small $\epsilon$ under the RG scaling.
This means that the timescales for the spinons are actually longer than for the gauge bosons near the IR fixed point
in this limit. This could 
suggest that the gauge bosons can be assumed, to a first approximation, to be in local thermal equilibrium (i.e., the
drag mechanism is a parametrically smaller effect in the system).
This point of view will be further elaborated later on in this article.
In such a physical regime where the drag effect can be neglected, 
we will show that the thermal conductivity is effectively dominated by the contribution of the spinons.
In this particular case, it is advantageous to keep $\eta$ initially finite and positive and set $\eta'=0$ from the outset.
We point out that the final answer is independent of $\eta'$.
Additionally, despite the fact that $\eta$ also flows to zero in the low-energy limit 
for $z_b>2$, we want to stress here that this prescription generates the right dynamics
for both spinons and gauge bosons since it reproduces correctly the Landau damping of the bosons
due to the gapless Fermi surface in the model \cite{Holstein,SSLee,Mross}. Indeed, the dressed 
fermion propagator and the boson propagator of the model become \cite{Mross}, respectively,

\vspace{-0.5cm}

\begin{eqnarray}
G_{f}^{(s)}(\mathbf{k},k_0)&=&\frac{1}{\frac{i}{\lambda N}\text{sgn}(k_0)|k_0|^{2/z_b}-\bar{\varepsilon}_{\mathbf{k},s}},\\
D_{b}(\mathbf{k},k_0)&=&\frac{1}{\frac{1}{4\pi}\frac{|k_0|}{|k_y|}+|k_y|^{z_b-1}},
\end{eqnarray}

\noindent where $\lambda=(4\pi)^{2/z_b}\sin(2\pi/z_b)$ and $\bar{\varepsilon}_{\mathbf{k},s}=s k_x+k_y^2$.

An important motivation for this work that we wish to emphasize is the fact that the conventional approach of deriving a quantum Boltzmann equation 
could potentially fail for this model because: (i) the quasiparticles are not well-defined
at low energies and (ii) a standard application of large-$N$ expansion for the
self-energy (similar to the Eliashberg approximation for an electron-phonon system \cite{Kadanoff}) leads to a strongly coupled theory \cite{SSLee}.
In order to circumvent some of these difficulties, we shall use in the present work the Mori-Zwanzig approach \cite{Woelfle,Forster} (also known
as the memory matrix formalism) which has the advantage of not relying on the quasiparticle picture. As will become clear shortly, 
the memory matrix can be viewed as a generalization
of the concept of scattering rate in Boltzmann theory applicable to systems
in which this quantity is not well-defined. Because of this appealing feature, the Mori-Zwanzig approach
has been successfully applied to one-dimensional interacting electrons \cite{Giamarchi,Rosch,Rosch2} 
and, in recent years, also to some higher dimensional systems
at quantum criticality \cite{Hartnoll2,Sachdev,Hartnoll}. In this formalism, only the operators with the 
longest relaxation timescales can potentially contribute to
the thermal conductivity of the system, since operators with short-time decay are in general irrelevant in the low-energy effective  
description. For this reason, these latter operators are not expected to play a key role in the calculation of the transport coefficients
in the present model.

Since the action $S$ is invariant under both space translation and global $U(1)$ symmetry, by following Noether's theorem
one finds that both the classical momentum and particle current densities are conserved.
These quantities at the quantum level will play a central role in our discussion, since one expects that their corresponding operators 
should have the longest relaxation timescales in the system and, for this reason, we will argue that they dominate the transport properties.
Accordingly, we shall write down the translation operator $\mathbf{P}_T$ in the system and the current operator $\mathbf{J}$, respectively, as follows: 
$\mathbf{P}_T=(P_{T}^x,P_{T}^y)=i\eta\int d^2 x \sum_{s\alpha}\nabla{\psi^{\dagger}}_{s\alpha}\psi_{s\alpha}
+2\eta'(\partial_t \phi)\nabla \phi$ and
$\mathbf{J}=(J_x,J_y)$, where $J_x=\int d^2 x \sum_{s\alpha} s\, v_x {\psi^{\dagger}_{s\alpha}}\psi_{s\alpha}$ and 
$J_y=i\sum_{s\alpha}v_y(\partial_y \psi^{\dagger}_{s\alpha}\psi_{s\alpha}-\psi^{\dagger}_{s\alpha}\partial_{y}\psi_{s\alpha})$ (see, e.g., Ref. \cite{Rosch} for a definition of 
the same physical quantities in the context of a one-dimensional (1D) model).
If all energies are measured with respect to the chemical
potential, the Hamiltonian density $h(\mathbf{x})$ of the model is the heat density.
Therefore, using the continuity equation for the heat flow $\dot{h}(\mathbf{x})+\nabla.\,\mathbf{J}_{Q}=0$ (where the dot stands for a time derivative), 
we can also formally obtain the thermal current operator $\mathbf{J}_Q=(J_{Q}^{x},J_{Q}^{y})$, 
whose components are given by

\vspace{-0.3cm}

\begin{align}
&{J_{Q}^{x}}=-\int d^2 x\sum_{s\alpha} \frac{is}{2}(\dot{\psi}^{\dagger}_{s\alpha}\psi_{s\alpha}-{\psi}^{\dagger}_{s\alpha}\dot{\psi}_{s\alpha}),
\label{thermal_1}\\
&{J_{Q}^{y}}=-\int d^2 x\sum_{s\alpha} \frac{is}{2}(\dot{\psi}^{\dagger}_{s\alpha}\partial_y\psi_{s\alpha}+\partial_y {\psi}^{\dagger}
_{s\alpha}\dot{\psi}_{s\alpha})\label{thermal_2}-(\partial_y \phi)^{\epsilon}\dot{\phi}.
\end{align}

\noindent To calculate the thermal conductivity of this model, it is important to include all the conserved (or nearly conserved) quantities 
in the system that have a finite 
overlap with the above thermal current operator. 

\begin{figure}[t]
 \centering
 \includegraphics[height=0.9in]{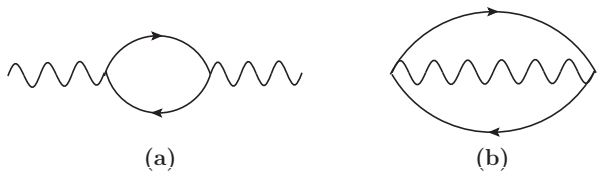}
 \caption{Generalized representation of the Feynman diagrams at leading order for finite $N$ and 
 small $\epsilon$ corresponding to the calculation of: (a) the matrix of susceptibilities $\hat{\chi}^{AB}$ and (b) the memory matrix $\hat{M}_{AB}$ 
 in the present model. The solid lines stand for the dressed fermion propagator, whereas the wavy lines
 represent the boson propagator. Note that the coupling parameters in both diagrams change depending on
 the operators and matrix elements being calculated.}
\end{figure}

To begin with and to set up our notation, we define a matrix of generalized conductivities in the following way

\vspace{-0.3cm}

\begin{eqnarray}
\hat{\sigma}(\omega,T)=\frac{i\hat{\chi}_{ret}(T)}{\omega+i\hat{M}(\omega,T)\hat{\chi}^{-1}_{ret}(T)}, 
\end{eqnarray}

\noindent where $\hat{\chi}_{ret}^{AB}(T)=\hat{\chi}_{ret}^{AB}(\omega=0,T)$ is the matrix of static retarded susceptibilities 
of some possible ``slowly-varying'' operators $A$ and $B$
in the system. This matrix of susceptibilities is defined in the conventional way by

\vspace{-0.3cm}

\begin{equation}
\hat{\chi}^{AB}(i\omega,T)=\int_{0}^{1/T}d\tau e^{i\omega\tau} \left\langle T_{\tau} A^{\dagger}(\tau) B(0)\right\rangle,
\end{equation}

\noindent where $\hat{\chi}_{ret}^{AB}(\omega)=\hat{\chi}^{AB}(i\omega\rightarrow \omega + i0^{+})$,
the statistical average $\left\langle ... \right\rangle$ is taken over the grand canonical ensemble, and the volume $V$ has been set
to unity. The memory matrix is then given by \cite{Forster}

\vspace{-0.3cm}

\begin{equation}\label{memory_matrix}
M_{AB}(\omega,T)=\int_{0}^{1/T}d\tau \left\langle \dot{A}^{\dagger}(0){Q}\frac{i}{\omega-{Q}L{Q}}{Q}\dot{B}(i\tau)\right\rangle.
\end{equation}

\noindent Here the ``super''-operator $L$ is the Liouville operator which is defined as $LA=[H,A]=-i\dot{A}$, where $H$ is the Hamiltonian, and $Q$ 
projects onto the space of operators perpendicular to all the slowly-varying operators $\{A,B,...\}$.
Here we have assumed that the slowly-varying operators in the model have the same signature under time reversal symmetry.
The memory matrix encodes the mechanism of relaxation of all the
nearly conserved quantities in the system. Due to the projection operator $Q$, this matrix is expected to be also a smooth function of the coupling
constant $g$ of the model and, for this reason, it can be evaluated in a perturbative way.

Consider first the slowly-varying operators in the present system which are given by $\{\mathcal{O}_i\}=\{\mathbf{P}_T,\mathbf{J}\}$. By
analyzing their equations of motion, we obtain that 

\vspace{-0.3cm}

\begin{align}
i\dot{\mathbf{P}}_T&=[\mathbf{P}_T,H]=\sum_{s\alpha}g_s\int d^2 x \psi^{\dagger}_{s\alpha}
\psi_{s\alpha}[\eta(\nabla a+i\mathbf{K}a)\nonumber\\
&-\eta'\nabla a)e^{i\mathbf{K}.\mathbf{x}}+..., \label{Eq9}\\
i\dot{J}_x&=[J_x, H]=0, \label{Eq10}\\
i\dot{J}_y&=[J_y, H]
=\sum_{s\alpha}g_s\int d^2 x \psi^{\dagger}_{s\alpha}
\psi_{s\alpha}(\partial_y a+i K_y a_y)\nonumber\\
&\times e^{i\mathbf{K}.\mathbf{x}}+...,\label{Eq11}
\end{align}

\noindent where the ellipses represent similar terms (but with no $\mathbf{K}$ contribution) generated by the normal part of the interaction between the spinons and
the gauge bosons. At this point, it is instructive to pause for a moment in order to analyze separately the contributions
arising from the $\eta$ and $\eta'$ terms. This is related to the discussion concerning a possible
existence (or non-existence) of the drag effect played by the gauge bosons in the present system. 
Since the bosons and spinons are weakly interacting in the limit of small $\epsilon$ and
finite $N$, it is in principle not clear whether they are able to exchange momentum efficiently
in such a case. This discussion is complicated by the fact that in real materials there
are always impurities or lattice defects -- which were not included in the present analysis --
that turn out to be quite effective in destroying
the excess momentum of the bosons, while having a negligible effect on the fermions \cite{Peierls,Ashcroft}.
In that case, the thermal conductivity of the spinons should follow a power-law temperature dependence that would have strong analogies
to Bloch's law for the electron-phonon scattering in metals.
In other words, it is indeed possible that the drag effect is masked by the scattering due to the above additional ingredients \cite{Peierls,Ashcroft}
and therefore we may assume, to a good approximation, that the gauge bosons are kept in equilibrium. For this regime,
we must set $\eta'$ to zero, since this is the most irrelevant
contribution near the IR nontrivial fixed point, as explained before in the present paper. 
This scenario will be analyzed in some detail below.
On the other hand, for samples of extremely high purity, the extrinsic mechanisms mentioned above
cannot be invoked any longer. In such a case, the drag effect of the gauge bosons
would be important in the system and, for this regime, we must set $\eta=\eta'$ in Eqs. (\ref{Eq9}), (\ref{Eq10}) and (\ref{Eq11}).
We will also analyze the impact of this drag effect on our results concerning the thermal conductivity
of the present system.  

Given the above discussion, let us assume henceforth the first scenario in which there is no drag effect of the gauge bosons in the system. 
Later on, we will also examine the second scenario.
The thermal conductivity can be formally written \cite{Ziman} 
as $\kappa=[\sigma_{\mathbf{J}_Q \mathbf{J}_Q}(\omega\rightarrow 0,T)]/T$, where

\vspace{-0.3cm}

\begin{equation}\label{kappa_equation}
\kappa=\frac{1}{T}\sum_{i,j}\chi_{{\mathbf{J}_Q}\mathcal{O}_i}(T)M_{\mathcal{O}_i \mathcal{O}_j}^{-1}(T)\chi_{\mathcal{O}_j {\mathbf{J}_Q}}(T).
\end{equation}

\noindent From Eq. (\ref{Eq10}) it appears at first sight that, since $J_x$ does not relax in the present model, 
the inverse of the memory matrix defined by Eq. (\ref{memory_matrix})
might display a divergence in our calculation, which would require invoking some extrinsic mechanism (i.e. not included in the present model) 
such as coupling to disorder, and phonons. 
Despite this remark, we point out that in fact the operator $\mathbf{J}$ will not contribute in an effective way to the 
thermal conductivity in the present model, since its $x$-component $J_x$ has no overlap with the thermal current operator 
(and its $y$-component is, up to a prefactor, 
equal to $y$-component of $\mathbf{P}_T$ for $\eta'=0$). In other words, in the limit of finite $N$ and small $\epsilon$, 
the susceptibility $\chi_{J_x J^Q_x}(T)$ (for $\mathbf{q}=0)$ given by the integral 

\vspace{-0.3cm}

\begin{align}
&\chi_{J_x J^Q_x}(T)=T\sum_{\substack{{s,\alpha} \\ {\omega_n}}}\int\frac{d^{2}\mathbf{k}}{(2\pi)^2}\frac{1}{\frac{i}{\lambda N}\text{sgn}(\omega_n)|\omega_n|
^{\frac{2}{z_b}}
-\bar{\varepsilon}_{\mathbf{k},s}}\nonumber\\
&\times\frac{(\omega_n + q_0/2)}{\frac{i}{\lambda N}\text{sgn}(\omega_n+q_0)|\omega_n+q_0|^{\frac{2}{z_b}}-\bar{\varepsilon}_{\mathbf{k},s}}
\end{align}

\noindent vanishes identically for any temperature $T$, i.e. $\chi_{J_x J^Q_x}(T)=0$ (see Fig. 2(a)
for a representation of the corresponding Feynman diagram in this calculation). 
The above integral can be computed straightforwardly by
using Cauchy's residue theorem and choosing a contour
of integration that avoids crossing two branch cuts that
exist in the complex plane.

In an analogous way, we can calculate the susceptibility $\chi_{P_T^x J^Q_x}(T)$ in the limit 
of finite $N$ and small $\epsilon$. Since $\eta$
will cancel out in the final result for the thermal conductivity, we set from this point on $\eta=1$
without loss of generality. 
In this case, we obtain
\vspace{-0.3cm}

\begin{align}
\chi_{P_T^x J^Q_x}(T)&=-\frac{4i}{\lambda N}\sum_{s,\alpha}s\int_{\mathbf{k}}\int_{\varepsilon} f(\varepsilon)\varepsilon\frac{|\varepsilon|^{2/z_b}
 \bar{\varepsilon}_{\mathbf{k},s}\,k_x}
{\left(\frac{|\varepsilon|^{4/z_b}}{\lambda^2 N^2}+\bar{\varepsilon}^2_{\mathbf{k},s}\right)^{2}},
\end{align}

\noindent where $\int_{\mathbf{k}}=\int \frac{d^2 \mathbf{k}}{(2\pi)^2}$, $\int_{\mathbf{\varepsilon}}=\int \frac{d\varepsilon}{(2\pi)}$
and $f(\varepsilon)=[1+e^{\beta\epsilon}]^{-1}$ is the Fermi-Dirac distribution.
It can be useful to rewrite the above integral in terms of the following dimensionless variables:
$\bar{x}=\beta^{2/z_b}k_x$, $\bar{y}=\beta^{1/z_b}k_y$, and $\bar{z}=\beta \varepsilon$, where $\beta=1/T$.
As a result, we obtain that this susceptibility scales as $\chi_{P_T^x J^Q_x}(T)= C_{z_b}^{(1)}(\lambda N)\, T^{5/2-\epsilon/4}$, where
the prefactor $C_{z_b}^{(1)}(\lambda N)$ is independent of $T$ and is finite for small $\epsilon$ and finite $N$ .
We perform a similar calculation for $\chi_{P_T^y J^Q_y}(T)$ and obtain that this quantity also
has the same temperature dependence as $\chi_{P_T^x J^Q_x}(T)$, i.e. $\chi_{P_T^y J^Q_y}(T)= C_{z_b}^{(2)}(\lambda N)\, T^{5/2-\epsilon/4}$,
but with a different temperature-independent prefactor $C_{z_b}^{(2)}(\lambda N)$.
One may then conclude from this analysis that the translation operator of spinons $\mathbf{P}_T$ 
will play the role of only slowly-varying operator in the thermal conductivity calculation for the present system.
Therefore, Eq. (\ref{kappa_equation}) can be further simplified to

\vspace{-0.3cm}

\begin{equation}\label{kappa/T}
\frac{\kappa}{T}=\frac{1}{T^2}\sum_{i,j=x,y}\chi_{J^i_Q P^i_T}(T)M_{P_T^i P_T^j}^{-1}(T)\chi_{P_T^j J_Q^j}(T).
\end{equation}

Next, we proceed to calculate the memory matrix operator $M_{P_T^i P_T^j}$ for the present model in a controlled way in the same limit as explained before.
We note that $\dot{\mathbf{P}}_T$ is of order linear in the coupling constant $g$. Therefore, from Eq. (\ref{memory_matrix}),
the dominant contribution to $\hat{M}$ is of order $O(g^2)$. Since we
want to keep only the leading order contribution to this quantity, we should set the Liouville operator
$L=L_0+gL_{int}$ to its noninteracting value ($L\approx L_0$). In addition,
following the same strategy, the grand-canonical average with the full Hamiltonian of the system must be also replaced by the corresponding average 
in the noninteracting limit, i.e. $\left\langle ... \right\rangle_{0}$. Lastly, since $L_0 P_T^i =0$ $(i=x,y)$, it can be 
shown that there is no contribution from the projection operator $Q$ in the present case, i.e. $L_0 Q=L_0$
and $Q\dot{P}^i_{T}=\dot{P}^i_{T}$ (see, e.g., Ref. \cite{Rosch} for
the same condition employed in the context
of a 1D model). Therefore, the memory matrix for the present $U(1)$ spin-liquid model becomes

\vspace{-0.3cm}

\begin{eqnarray}\label{memory_matrix_2}
{M}_{P_T^iP_T^j}(\omega\rightarrow 0,T)&\approx&\lim_{\omega\rightarrow 0}\frac{[\langle\dot{{P}}_T^i;\dot{P}_T^j\rangle_{\omega}^{0}
-\langle\dot{P}_T^i;\dot{P}_T^j\rangle_{\omega=0}^{0}]}{i\omega}\nonumber\\
&=&-i\frac{\partial}{\partial\omega}\langle\dot{P}_T^i;\dot{P}_T^j\rangle_{\omega}^{0}\big|_{\omega=0},
\end{eqnarray}

\noindent where $\langle\dot{P}_T^i;\dot{P}_T^j\rangle_{\omega}^{0}=\langle\dot{P}_T^i(\omega)\dot{P}_T^j(-\omega)\rangle_{0}$.
Let us concentrate first on the term $\langle\dot{P}_T^x;\dot{P}_T^x\rangle_{\omega}^{0}$. 

Using Wick's theorem (see Fig. 2(b) for a representation of the corresponding Feynman diagram), its leading term is given by 

\vspace{-0.3cm}

\begin{align}\label{memory_matrix_3}
&\langle\dot{P}_T^x(\omega)\dot{P}_T^x(-\omega)\rangle_{0}=-T^2\sum_{\substack{{s,\alpha} \\ {k_0,k'_0}}}g^2\int_{\mathbf{k,k'}}
G_{f}^{(s)}(\mathbf{k'+ K},k'_0)\nonumber\\
&\times G_{f}^{(s)}(\mathbf{k},k_0) D_{b}(\mathbf{k'}-\mathbf{k},k'_0-k_0+\omega)(k'_x-k_x)^2 +\ldots,
\end{align}

\noindent where the ellipsis refers to the subleading terms. Using Cauchy's residue theorem
with a choice of contour of integration that
circumvents crossing a branch cut that exists in the complex plane,
the leading term of Eq. (\ref{memory_matrix_3}) can be written as

\vspace{-0.4cm}

\begin{align}\label{memory_matrix_2}
&\langle\dot{P}_T^x(\omega)\dot{P}_T^x(-\omega)\rangle_{0}=\frac{4}{\lambda^2 N^2}\sum_{s,\alpha}g^2\int_{\mathbf{k,k'}}
\int_{\varepsilon,\varepsilon'}\frac{f(\varepsilon)f(\varepsilon')}{\left(\frac{|\varepsilon|^{4/z_b}}{\lambda^2 N^2}+
\overline{\varepsilon}_{\mathbf{k},s}^2\right)}\nonumber
\\&\times\frac{|\varepsilon|^{2/z_b} |\varepsilon'|^{2/z_b}(k'_x-k_x)^2}{\left(\frac{|\varepsilon'|^{4/z_b}}{\lambda^2 N^2}+
\overline{\varepsilon}_{\mathbf{k'+ K},s}^2\right)\left(\frac{1}{4\pi}\frac{|\varepsilon'-\varepsilon+i\omega|}{|k'_y-k_y|}+|k'_y-k_y|^{z_b-1}\right)}+\ldots
\end{align}

\noindent It can also be advantageous here to rewrite the above integral in terms of the following dimensionless variables:
$\bar{x}=\beta^{2/z_b}k_x$, $\bar{x}'=\beta^{2/z_b}k'_x$, $\bar{y}=\beta^{1/z_b}k_y$, $\bar{y}'=\beta^{1/z_b}k'_y$,
$\bar{z}=\beta \varepsilon$ and $\bar{z}'=\beta \varepsilon'$.
Then, by taking the derivative with respect
to $\omega$, we obtain that $M_{P_T^xP_T^x}(T)=C_{z_b}^{(3)}(\lambda N)\,T^{\frac{7}{4}(2-\epsilon)}$ for intermediate temperatures, where
the new prefactor $C_{z_b}^{(3)}(\lambda N)$ is again independent of $T$ and finite in the limit of small $\epsilon$ and finite $N$.
Moreover, we can follow the same strategy to calculate $\langle\dot{P}_T^y;\dot{P}_T^y\rangle_{\omega}^{0}$. Indeed, in an analogous way to the 
previous computation \cite{Remark}, we obtain in the same limit as before that $M_{P_T^yP_T^y}(T)=C_{z_b}^{(4)}(\lambda N)\, T^{\frac{5}{4}(2-\epsilon)}$
(with a prefactor $C_{z_b}^{(4)}(\lambda N)$ also independent of $T$).

The last step consists of analyzing the off-diagonal terms in the memory matrix, i.e. ${M}_{P_T^xP_T^y}$ and ${M}_{P_T^yP_T^x}$.
By performing a similar calculation as outlined above, the leading terms can be shown to vanish in the same limit. 
Therefore, the memory matrix turns out to be approximately diagonal in the basis 
$\{P_T^x,P_T^y\}$ and can
be very easily inverted. Indeed

\vspace{-0.5cm}

\begin{equation}
\hat{M}^{-1}\approx\left( \begin{array}{cc}
M^{-1}_{P_T^xP_T^x} & 0 \\
0 & M^{-1}_{P_T^yP_T^y} \\
\end{array} \right).
\end{equation}

\noindent As a result, by means of Eq. (\ref{kappa/T}), we obtain that the thermal conductivity $\kappa$ is given by

\vspace{-0.3cm}

\begin{equation}
\frac{\kappa}{T}\approx\frac{1}{T^2}\left[\chi_{P_T^x J^Q_x}^2(T)M^{-1}_{P_T^xP_T^x}(T)+
\chi_{P_T^y J^Q_y}^2(T)M^{-1}_{P_T^yP_T^y}(T)\right],
\end{equation}

\noindent which yields

\vspace{-0.3cm}

\begin{equation}\label{kappa}
\frac{\kappa}{T}\approx A_{xx}\,T^{-\frac{1}{2}+\frac{5\epsilon}{4}}+
A_{yy}\,T^{+\frac{1}{2}+\frac{3\epsilon}{4}},
\end{equation}

\noindent with finite prefactors $A_{xx}={[C_{z_b}^{(1)}(\lambda N)]^2}/{C_{z_b}^{(3)}(\lambda N)}$ and 
$A_{yy}={[C_{z_b}^{(2)}(\lambda N)]^2}/{C_{z_b}^{(4)}(\lambda N)}$. 
In the limit of small $\epsilon$, one can verify that the second term on the rhs in Eq. (\ref{kappa}) 
is subleading. Therefore, assuming the first scenario in which there is no drag effect
of the gauge bosons in the system, there will be a regime in which
the first term on the rhs of Eq. (\ref{kappa}) will dominate over the second term at intermediate temperatures. In other words,
we obtain in this work that the thermal conductivity $\kappa$ of the present model should
scale within a certain temperature regime as a power-law $\kappa\sim T^{\gamma}$ with the exponent 
$\gamma=1/2+5\epsilon/4$, due to non-critical Umklapp
scattering of the spinons
for a finite number of fermionic species $N$ and a small parameter $\epsilon=z_b -2$. 
As a consequence, we can infer from this result that
a possible tendency towards a linear behavior as a function of $T$ in the thermal conductivity of this model with
$\kappa\sim T^{\gamma}$ for $1/2<\gamma\leq 1$ is indicated qualitatively by our approach
at those temperatures. Despite this, we point out that a quantitative
comparison with the experimental situation (e.g., in the organic material EtMe$_3$Sb[Pd(dmit)$_2$]$_2$ \cite{Yamashita}) 
still cannot be rigorously established. This is 
related to the fact that the IR nontrivial fixed point coupling obtained in this problem for $\epsilon\rightarrow 1$
(and finite $N>1$) becomes strong,
which would require the perturbative calculation of the memory matrix and the susceptibilities beyond the
lowest order considered in the present work. Such a more complicated analysis would be valuable in order to carry out a precise
comparison of the present theoretical prediction with the experimental situation.

Next, we move on to discuss the second scenario in which there could be the drag effect associated
with the gauge bosons and, consequently, those excitations are never in equilibrium in the system. In this case,
as was explained before, we must set $\eta=\eta'$ in Eqs. (\ref{Eq9}), (\ref{Eq10}) and (\ref{Eq11}).
The rest of the calculation proceeds essentially in a similar way as was performed for the previous
scenario, with only minor modifications. Most importantly, one of the leading terms consisting of $\langle\dot{P}_T^x(\omega)\dot{P}_T^x(-\omega)\rangle_{0}$
discussed previously becomes for the present case  

\vspace{-0.4cm}

\begin{align}\label{memory_matrix_2}
&\langle\dot{P}_T^x(\omega)\dot{P}_T^x(-\omega)\rangle_{0}=\frac{4}{\lambda^2 N^2}\sum_{s,\alpha}g^2\int_{\mathbf{k,k'}}
\int_{\varepsilon,\varepsilon'}\frac{f(\varepsilon)f(\varepsilon')}{\left(\frac{|\varepsilon|^{4/z_b}}{\lambda^2 N^2}+
\overline{\varepsilon}_{\mathbf{k},s}^2\right)}\nonumber
\\&\times\frac{|\varepsilon|^{2/z_b} |\varepsilon'|^{2/z_b}\,(iK_x)^2}{\left(\frac{|\varepsilon'|^{4/z_b}}{\lambda^2 N^2}+
\overline{\varepsilon}_{\mathbf{k'+ K},s}^2\right)\left(\frac{1}{4\pi}\frac{|\varepsilon'-\varepsilon+i\omega|}{|k'_y-k_y|}+|k'_y-k_y|^{z_b-1}\right)},
\end{align}

\noindent from which follows that the thermal conductivity of the system in this second scenario
would be modified to $\kappa\sim T^{\gamma'}$ with the exponent being $\gamma'=5/2+\epsilon/4$. Since this latter theoretical prediction
is not observed experimentally, this may suggest that the first scenario appears to be more appropriate
to describe the transport properties of the candidate spin-liquid phase of the organic material EtMe$_3$Sb[Pd(dmit)$_2$]$_2$
in the literature \cite{Yamashita}.

Another important point we want to emphasize here is that, at very low temperatures, Umklapp processes turn out to be exponentially suppressed as a function
of temperature and, in this regime, they cannot provide any longer an efficient mechanism to degrade the total momentum of the spinons in the present model.
As a result, the thermal conductivity of the system in the clean limit is eventually expected to
become exponentially enhanced at very low temperatures. 
We thus conclude that, in this low-$T$ regime, extrinsic mechanisms for momentum relaxation (such as, coupling to disorder) 
must be taken into account to produce a finite heat current in the system. If the disorder is extremely weak, Umklapp
processes dominate the transport properties of the system for a reasonably wide range of experimentally measured temperatures,
as obtained in the present work. Despite this, one can verify 
from dimensional
analysis that coupling the spinons (and also the gauge bosons) to impurities in the present model always represents a relevant perturbation
near the IR nontrivial fixed point. 
Therefore, for very low temperatures, a crossover of $\kappa$ to a different temperature dependence 
is expected. We leave open this problem here, since it is beyond the scope of the present work. 
We plan to perform a detailed analysis of the role played by disorder in the present system in a future publication.

Lastly, we point out that the field theory model defined by Eq. (\ref{action}) has some similarities with the theory associated with quantum criticality near 
a Pomeranchuk phase transition. Indeed, in Eq. (\ref{action}), if we alter the condition
for the coupling constant in the model from $g_{s}=sg$ (where $s=\pm$) to $g_s=g$, the resulting theory will describe instead an
Ising-nematic transition out of a metallic state which breaks the point-group rotation symmetry of the lattice 
but preserves translational symmetry \cite{Metlitski,Mross,Kivelson2,Metzner,Pepin,Metlitski2}.
It would be very interesting to apply similar analytically controlled methods to calculate the
thermal conductivity of such quantum critical metals
in the presence of both Umklapp and disorder to treat the competition between those two effects. 
There is an ongoing effort in this direction using many
techniques by several groups (see, e.g., Refs. \cite{Chubukov,Hartnoll,Raghu,SSLee3,Hartnoll2}).
In an important recent work, Ref. \cite{Hartnoll2} provided a thorough analysis of the impurity effects in
the calculation of the resistivity for this latter problem using the memory matrix formalism.
As a result, they pointed out that it is crucial to treat the effects of random-field disorder
on the order parameter in such a system at very low temperatures in a 
completely nonperturbative way.
In this respect, it goes without saying that comparing all the results regarding the transport coefficients
of the different models described here in this work
with other analytically controlled approaches to those problems based on, e.g., holographic methods would also be extremely helpful.

\begin{acknowledgments}
I would like to thank Sung-Sik Lee and, especially, T. Senthil for very useful discussions. 
I acknowledge financial support from CNPq under Grant No. 245919/2012-0 for this project.
\end{acknowledgments}


\begin{thebibliography}{99}
\bibitem{Remark} In fact, one has to be careful in this case because of the gauge boson contribution in Eq. (5).
Nevertheless, this term gives a subleading contribution that vanishes in the limit of small $\epsilon$ and finite $N$.
\end{thebibliography}
\end{document}